\RequirePackage{fix-cm}

\documentclass[smallextended]{svjour3}

\usepackage{graphicx}
\usepackage[caption=false]{subfig}
\usepackage{natbib}

\usepackage{url}

\usepackage{xspace}
\makeatletter
\xspaceaddexceptions{\check@icr}
\makeatother

\usepackage{ifpdf}
\ifpdf
       \usepackage[pdftex,colorlinks,pdfauthor={Hugo Buddelmeijer}]{hyperref}
       \DeclareGraphicsExtensions{.pdf,.png,.jpg}
\else
       \usepackage[hypertex,colorlinks]{hyperref}
\fi

\newcommand{\sourcecollection}{Source Collection\xspace}

\newcommand{\sourcecollections}{Source Collections\xspace}

\newcommand{\SourceCollection}{Source Collection\xspace}
\newcommand{\SourceCollections}{Source Collections\xspace}

\newcommand{\AW}{{\sf Astro-WISE}\xspace}
\newcommand{\aw}{{\sf Astro-WISE}\xspace}

\newcommand{\aweprompt}{{\tt awe}-prompt\xspace}

\newcommand{\topcat}{Topcat\xspace}
\newcommand{\aladin}{Aladin\xspace}

\newcommand{\samp}{SAMP\xspace}

\newcommand{\samphub}{HUB\xspace}

\newcommand{\refsec}[1]{section~\ref{#1}\xspace}
\newcommand{\reffig}[1]{Fig.~\ref{#1}\xspace}

\newcommand{\Reffig}[1]{Fig.~\ref{#1}\xspace}

\newcommand{\awfunction}[1]{\texttt{#1()}\xspace}

\newcommand{\treeviewer}{Tree Viewer\xspace}
\newcommand{\TreeViewer}{Tree Viewer\xspace}
\newcommand{\simplepuller}{Simple Puller\xspace}
\newcommand{\SimplePuller}{Simple Puller\xspace}
\newcommand{\objectviewer}{Object Viewer\xspace}
\newcommand{\ObjectViewer}{Object Viewer\xspace}

\newcommand{\sampcatalogderive}{target.\-catalog.\-derive\xspace}
\newcommand{\sampcatalogpull}{target.\-catalog.\-pull\xspace}
\newcommand{\sampobjecthighlight}{target.ob\-ject.\-highlight\xspace}
\newcommand{\sampobjectinfo}{target.ob\-ject.\-info\xspace}
\newcommand{\sampobjectchange}{target.ob\-ject.\-change\xspace}
\newcommand{\sampobjectaction}{target.ob\-ject.\-action\xspace}

\newcommand{\sampcataloghighlight}{\sampobjecthighlight}
\newcommand{\sampcataloginfo}{\sampobjectinfo}
\newcommand{\sampcatalogchange}{\sampobjectchange}
\newcommand{\sampcatalogaction}{\sampobjectaction}

\newcommand{\sampperscatalog}{catalog\xspace}

\newcommand{\asamppersobject}{an object\xspace}

\newcommand{\statusok}{ok}
\newcommand{\statusauto}{automatic}
\newcommand{\statusnew}{new}
\newcommand{\statusdepends}{depends}
\newcommand{\statusnot}{not}
\newcommand{\statusunknown}{unknown}

\begin{document}

\title{Query Driven Visualization of Astronomical Catalogs}


\author{
Hugo Buddelmeijer
           \and
 Edwin A. Valentijn
}


\institute{
Hugo Buddelmeijer \at
Kapteyn Astronomical Institute, Postbus 800, 9747 AD, Groningen, The Netherlands \\
\email{buddel@astro.rug.nl}           
           \and
 Edwin A. Valentijn \at
\email{valentyn@astro.rug.nl}           
}

\date{Received: date / Accepted: date}

\maketitle

\begin{abstract}
Interactive visualization of astronomical catalogs requires novel techniques due to the huge volumes and complex structure of the data produced by existing and upcoming astronomical surveys.
The creation as well as the disclosure of the catalogs can be handled by data pulling mechanisms \citep{pullingcatalogs}.
These prevent unnecessary processing and facilitate data sharing by having users request the desired end products.

In this work we present query driven visualization as a logical continuation of data pulling.
Scientists can request catalogs in a declarative way and set process parameters directly from within the visualization.
This results in profound interoperation between software with a high level of abstraction.

New messages for the Simple Application Messaging Protocol are proposed to achieve this abstraction.
Support for these messages are implemented in the \AW information system and in a set of demonstrational applications.

\end{abstract}

\section{Introduction}
\label{sec:qdvintro}
Large astronomical surveys require novel ways for handling the data they produce.
For example, the ongoing KiDS and VIKING surveys will cover 1500 square degree in optical and infra-red wavelengths \citep{2007Msngr.127...28A} and the upcoming Euclid mission will cover 20\thinspace000 square degree \citep{2009arXiv0912.0914L}.
These surveys will detect billions of galaxies for which hundreds of parameters will be quantified, leading to terabytes of data to explore.

\textit{Data pulling} mechanisms can be used to achieve the scalability to create catalogs (\citet{pullingcatalogs}, hereafter Paper I).
The essence of data pulling is that processing steps necessary to create a catalog are determined by specifying the required target catalog.
The information system will determine how existing catalogs can be used to fulfill the request and will initiate the creation of new catalogs only when no suitable ones exist.
This maximizes reusability of the catalogs and minimizes unnecessary calculations.
This requires full \textit{data lineage}, which means that catalogs are stored with all the information required to process them.

\textit{Query driven visualization} is a methodology to explore large data sets by limiting the processing required for visualization to the subsets of the data deemed ``interesting'' as defined by the user \citep{Stockinger:2006:DDS:1188455.1188542}.
Related work focuses on limiting the processing of the visualization itself \citep{Stockinger:2006:DDS:1188455.1188542}, the fast identification and retrieval of data \citep{10.1109/VIS.2005.84,4035755} or on the data representation \citep{4015432}.

In this paper, we see query driven visualization as the logical continuation of data pulling in an information system with full data lineage.
The main contributions of our work follow from applying this novel viewpoint to source catalogs: 
 (1) We limit the processing required to create the requested catalog itself, instead of the processing required for the visualization.
 (2) We permit requests in a more declarative form than direct database queries would allow.
 (3) We allow the user to inspect and influence the processing from within the visualization by exporting the data lineage.
 (4) We achieve a high level of abstraction that allows close interoperation between software.

We demonstrate our techniques with our \AW implementation and by designing new messages for the Simple Application Messaging Protocol.

\subsection{Data Pulling and Declarative Querying}
Data pulling is an excellent opportunity for query driven visualization. 
The autonomous discovery and creation of catalogs permits requests that are very declarative.
A scientist can request parameters of sources without having knowledge of whether these parameters have already been calculated or not.
This functionality can be implemented in external software and an example program to pull catalogs is given with the `\simplepuller' of \refsec{sec:qdvproto}.

Compare this for example with an SQL-based system \citep{Codd:1970:RMD:362384.362685}: to formulate an SQL query it is required to know which tables contain the required parameters, how to identify the relevant rows and columns, and often how to join tables.
This becomes a non-trivial problem when catalogs are shared between multiple users and the number of catalogs and their sizes grow.
At a certain stage it becomes too time consuming and error-prone to find required data by hand, especially when it is unknown whether it exists at all.

\subsection{Full Data Lineage and Exploration}
An information system with data pulling often has \textit{persistent objects} with \textit{full data lineage}:
Each data set is represented as an object---in computer science terminology---that persists between sessions and users.
In \AW these objects are called \textit{process targets}.
A process target contains all the information required to create the data it represents from other process targets, its \textit{dependencies}.
This is called \textit{backward chaining} and links every data project back to the raw data.

The data lineage can be utilized in query driven visualization by having the visualization software request it.
This allows the visualization software to show this information, either directly or processed in the  visualization.
An example of the former is given with the `\treeviewer' in \refsec{sec:qdvproto}.

Furthermore, exporting the data lineage makes it possible for scientists to influence the processing by permitting the visualization software to change processing parameters.
An example of this is given with the `\objectviewer' in \refsec{sec:qdvproto}.

\subsection{Abstraction and Interoperation through SAMP}
Data pulling mechanisms are well suited for abstraction on different levels: 
firstly, pulling data does not require detailed knowledge of every processing step;
secondly, these processing details themselves can be abstracted, because of the standardized data lineage.

Such an abstraction allows query driven visualization to be performed between any visualization package and information system.
The thoroughness of the interoperation will depend on the level of abstraction supported by both applications.
We extended the Simple Application Messaging Protocol\footnote{\url{http://www.ivoa.net/Documents/latest/SAMP.html}} to facilitate such interoperation by designing new message types (\refsec{sec:samp}).

\subsection[Astro-WISE]{\AW}
Query driven visualization requires an information system responsible for creation, storage and delivery of the data.
We choose to use \AW for this, although any information system with data pulling and persistent objects would be suitable, because of the abstraction through SAMP.
In \refsec{sec:sampastrowise} we describe the details of our \AW SAMP implementation.

\section{Interoperability through SAMP}
\label{sec:samp}
The \textit{Simple Application Messaging Protocol} (SAMP) is an International Virtual Observatory Alliance (IVOA) standard for interoperation between astronomical software.
The idea behind it is akin to the UNIX-philosophy that tools should do one thing, should do that thing well and communicate with other programs for things they cannot do.

\subsection{Simple Application Messaging Protocol}
We give a short description of SAMP before discussing our additions.
For details we refer to \refsec{sec:sampprotocol} and to the official documentation\footnote{\url{http://www.ivoa.net/Documents/latest/SAMP.html}}.
The protocol uses a client-server model based on application defined messages.
Clients register with the SAMP \samphub and subscribe to certain types of messages.
Clients can then send messages to individual clients or to any client that has registered for that kind of message.
The receiving application should subsequently perform the action it has associated with the message.
Lastly, the \samphub will relay a response back to the sender if necessary.

The expected action that corresponds to a message is determined by the type of the message.
Both default administrative messages and widely accepted application defined messages can be found on the SAMP wiki\footnote{\url{http://www.ivoa.net/cgi-bin/twiki/bin/view/IVOA/SampMTypes}}.
In the rest of this section we first describe relevant existing messages and subsequently introduce our proposed messages.
We list the type of the messages and a description of the intended action of the receiver.
Details of the messages and their parameters are given \refsec{sec:sampprotocolmessages}.

\subsection{Existing Catalog Related Messages}
Several existing catalog related messages can be used in conjunction with our new messages:
\begin{itemize}
\item \texttt{table.load.votable}: Load a table in VOTable format.
\item \texttt{table.load.fits}: Load a table in FITS format.
\item \texttt{table.highlight.row}: Highlight a single row of an identified table.
\item \texttt{table.select.rowList}: Select a list of rows of an identified table.
\end{itemize}
Exactly what `highlighting' or `selecting' means is left to the receiving application.
Tables have three identifiers in SAMP: a table-id that is unique within the SAMP session, a URI where the catalog can be found and a human readable name.
These identifiers are set with one of the \texttt{load} messages and used as a reference in the other messages. 
Rows are identified by their position in the table using zero-based indexing.
Note that these messages can refer to any tabulated data set. 
In this paper we limit ourselves to source catalogs only.

\subsection{Data Pulling Messages}
\label{sec:sampqdv}
We designed new SAMP messages to create a system independent way to perform pulling of catalog data.
The messages should be sent from visualization software to the information system handling the data.
The new message types start with \texttt{target.}; this is the name that the \AW information system uses to describe data objects that can be pulled:
\begin{itemize}
 \item \texttt{\sampcatalogpull}: Pull a catalog and send it over SAMP using one of the \texttt{table.\-load.*} messages.
The result could be an existing catalog or a new catalog created by the pulling mechanisms.
Any new data that is necessary to produce the required catalog is created automatically.
This message requires the following parameters, detailed below: an identifier of a catalog to select the sources from, a selection criterion and a list of requested attributes of the sources.
 \item \texttt{\sampcatalogderive}: Derive a catalog in the same fashion as with \texttt{\sampcatalogpull}, but do not create any new data or send the catalog data over SAMP.
\end{itemize}
Support for the \texttt{.pull} message is the minimum required to request catalog data from the information system.
The \texttt{.derive} message is useful when it is necessary to inspect or modify the derivation of the catalog---using the messages in \refsec{sec:samppersobj}---before visualization, for example to determine whether all required data is processed already or whether new data has to be created.
These two messages require three parameters which we should elaborate on (see also \refsec{sec:sampprotocolqdv}):
\begin{itemize}
 \item \texttt{catalog-id}: An identifier of the base catalog to select the sources from.
It is left to the information system to inform scientists how to refer to a specific catalog.
The \texttt{catalog-id} can be a unique identifier of an existing catalog, but could also be a reference to a catalog that does not yet exist, e.g. a photometric catalog for an observation that has not yet been reduced.
It is also possible to designate identifiers for special catalogs, e.g. to denote the latest version of a catalog of an ongoing survey. 
 \item \texttt{query}: 
A selection criterion to specify which sources of the original catalog are requested.
This should be a logical expression referencing the \texttt{attributes} below.
The exact specification of this expression is left to the information system.
A logical choice would be the syntax of an \texttt{ADQL}\footnote{\url{http://www.ivoa.net/Documents/latest/ADQL.html}} \texttt{WHERE} clause (without the `\texttt{WHERE}' itself).
 \item \texttt{attributes}:
A list of requested attributes (parameters) of the sources.
It is not required that the catalog corresponding to the \texttt{catalog-id} contains these attributes.
The data pulling mechanisms of the information system should try to find the requested attributes in related catalogs and should create new data sets if necessary.
How an attribute should be specified, is left to the information system.
\end{itemize}

\subsection{Object Messages}
\label{sec:samppersobj}
Several SAMP message types are defined for interaction with an information system with persistent objects.
These messages allow the visualization software to gain information about the objects and inspect or even influence its processing.
The persistent object related messages are:
\begin{itemize}
 \item \texttt{\sampcataloghighlight}: Highlight an object.
 \item \texttt{\sampcataloginfo}: Return information about an object, see below.
 \item \texttt{\sampcatalogchange}: Change the value of a property of an object such as a process parameter or a dependency.
 \item \texttt{\sampcatalogaction}: Perform an action related to an object or property. Possible actions are retrieved using the \texttt{\sampcataloginfo} message.
\end{itemize}
The \texttt{\sampcataloghighlight} message can be sent to any application, the others are supposed to be sent to the information system only.

A specific SAMP map is defined as a return value for the \texttt{\sampcataloginfo} message, containing information about the object and its properties (see \refsec{sec:sampprotocol} for details).
For the object itself it includes information about what properties it has, its processing status and whether the object can be modified.

The properties of an object include process parameters and references to the dependencies of the object.
The returned information about a property include its name, current value and optionally other values it can be set to.
Furthermore the information system can define actions that can be performed on the object or its properties.

\section{SAMP HUB and Clients}
\begin{figure}[ht!]
 \centering
 \includegraphics[width=0.9\linewidth]{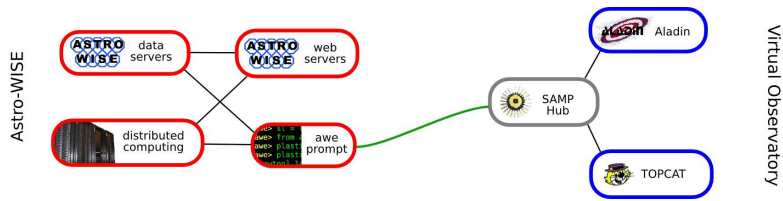}
 \caption{The connectivity between \aw\ and \samp. The \samp\ HUB in the center, the \aw\ system on the left and other \samp\ enabled applications on the right.
}
\label{fig:awesamp}
\end{figure}

The new SAMP messages are implemented in the \AW information system and demonstrated by a set of proof-of-concept applications.
We first describe relevant existing SAMP applications, subsequently the \AW\ \samp connectivity and end with the applications to demonstrate the new messages. 
\Reffig{fig:awesamp} shows a diagram of the interoperability between \aw\ and several \samp applications.

\subsection{Existing SAMP applications}
We list existing SAMP applications that are relevant to catalog data.

\begin{itemize}
\item \textbf{SAMP HUB}:
The \textit{HUB} is the center of SAMP to which the other applications connect. 
The HUB can be a standalone program or can be integrated in one of the clients, e.g. \aladin\ and \topcat include one.

\item \textbf{Topcat}:
\textit{\topcat}\footnote{\url{http://www.starlink.ac.uk/topcat/}} is a table viewer/manipulator written in Java.
The visualization power of \topcat\ lies in its interactivity.
Selections performed in one window propagate to other windows and by the use of SAMP messages to other applications.

\item \textbf{Aladin}:
\textit{\aladin}\footnote{\url{http://aladin.u-strasbg.fr/}} 
is an interactive software sky atlas allowing the user to visualize digitized astronomical images up to 50K by 50K pixels, superimpose entries from astronomical catalogues or databases, and interactively access related data and information online archives for all known sources in the field.

\end{itemize}

\subsection[Astro-WISE and SAMP]{\AW and SAMP}
\AW has SAMP connectivity in the interactive Python prompt and on the webservices. 

\begin{itemize}
\item \textbf{\aweprompt}:
The \AW\ \aweprompt is an interactive Python prompt that forms the primary user interface to \AW.
We developed a module for SAMP connectivity in the \aweprompt and other Python applications.
All messages from section \ref{sec:samp} are supported.

\item \textbf{DBViewer}:
With the \AW\ \textit{DBViewer} one can view all content of the database and can send query results over SAMP.
The DBViewer is beyond the scope of this paper.
\end{itemize}

\subsection{Query Driven Visualization Prototype}
\label{sec:qdvproto}

A set of proof-of-concept applications has been developed to demonstrate different ways in which SAMP clients can use the query driven visualization messages.
They interact with the \AW\ \aweprompt through SAMP only and have little knowledge about \AW, if at all.

\begin{figure}[htp]
 \centering
 \includegraphics[width=0.5\linewidth]{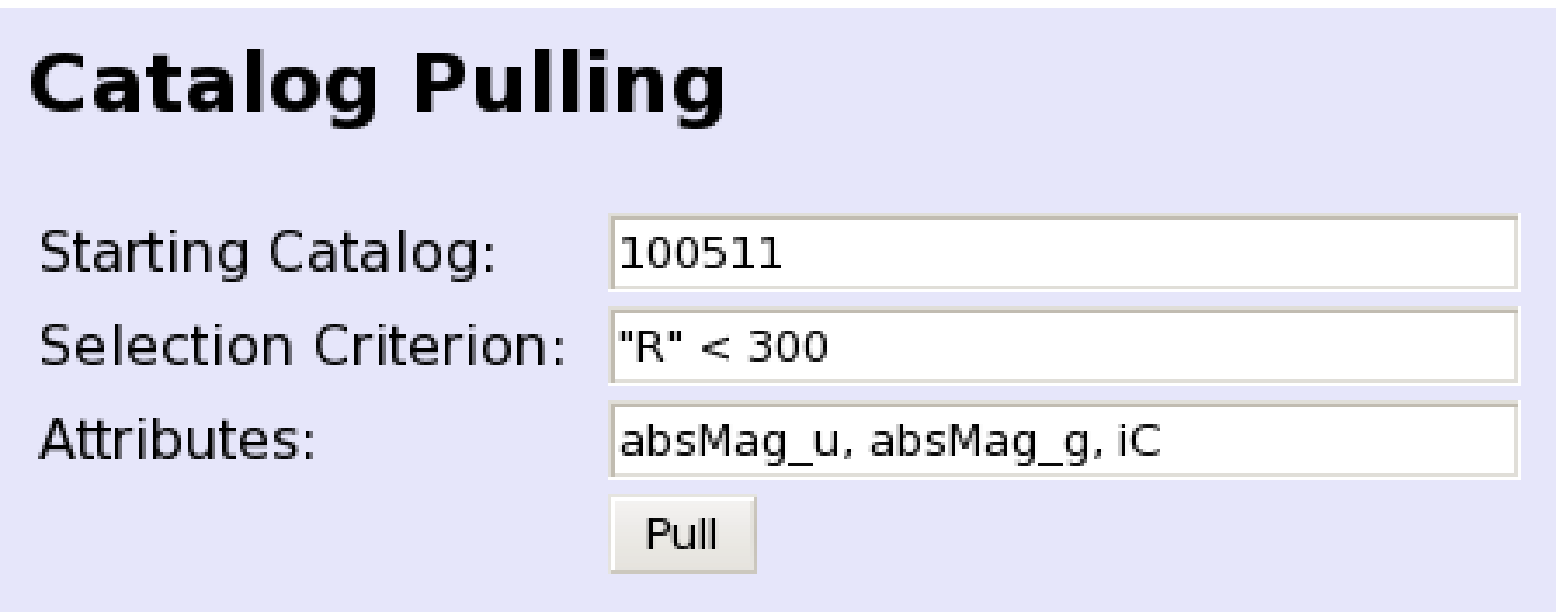}
 \caption{The \SimplePuller application for pulling catalogs over SAMP.
It can pull data from any information system that accepts the \texttt{\sampcatalogpull} message.
}
 \label{fig:scsimplepuller}
\end{figure}

\begin{figure}[htp]
 \centering
 \includegraphics[width=0.7\linewidth]{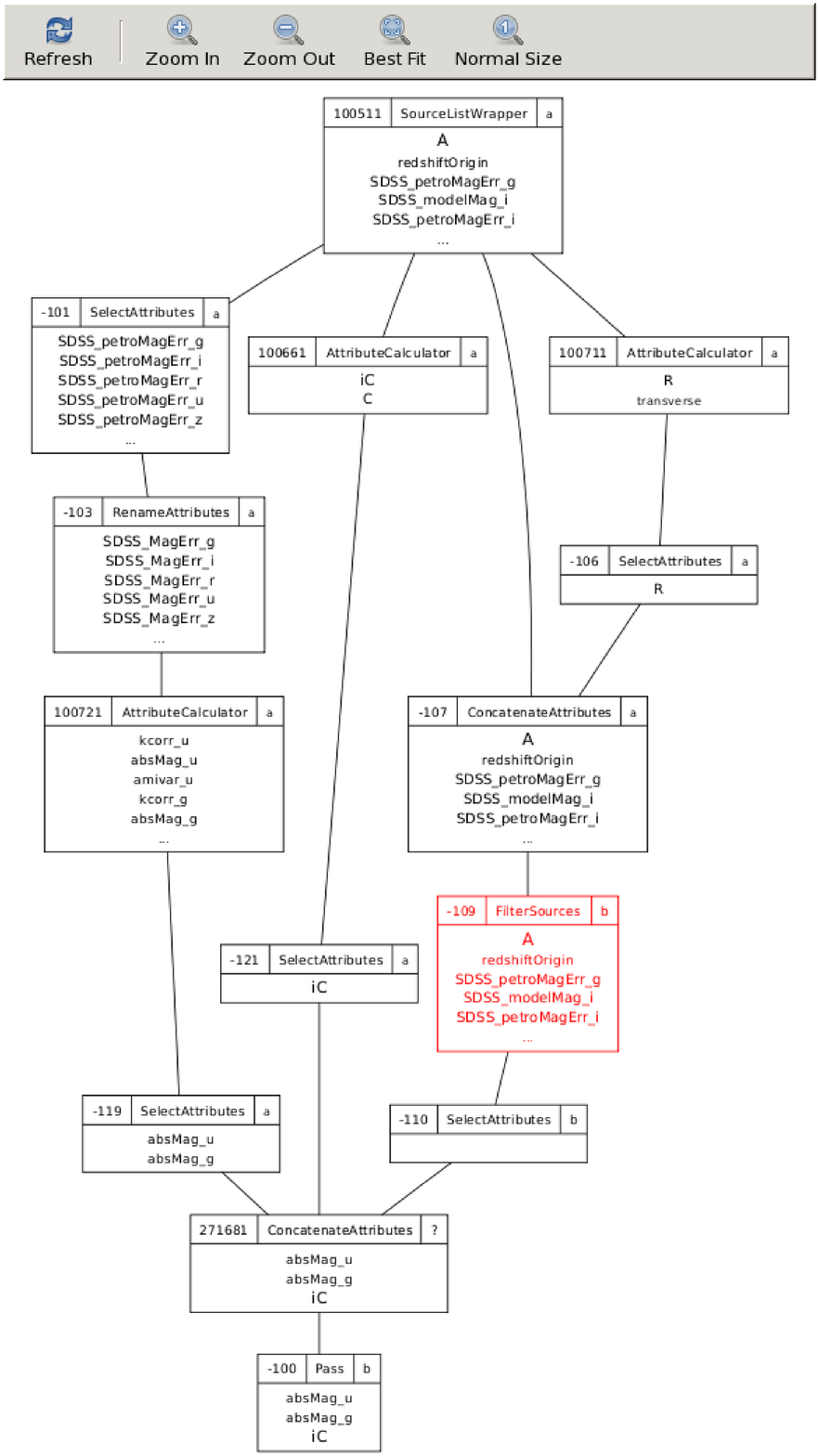}
 \caption{SAMP application for exploring dependency graphs of catalog objects in \AW.
Every node shows the catalog identifier on the top left, the class of the catalog in the top center and an
identifier for the set of sources on the top right.
The attributes of the sources are shown in the rest of the box.
}
 \label{fig:sctreeexplorer}
\end{figure}

\begin{figure}[htp]
 \centering
 \includegraphics[width=0.5\linewidth]{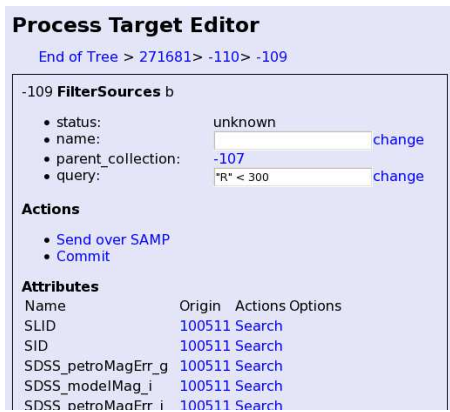}
 \caption{
SAMP application to view and modify details of individual catalogs or other objects.
The highlighted catalog from \reffig{fig:sctreeexplorer} is shown.
}
 \label{fig:scscviewer}
\end{figure}

\begin{figure}[htp]
 \centering
 \includegraphics[width=0.5\linewidth]{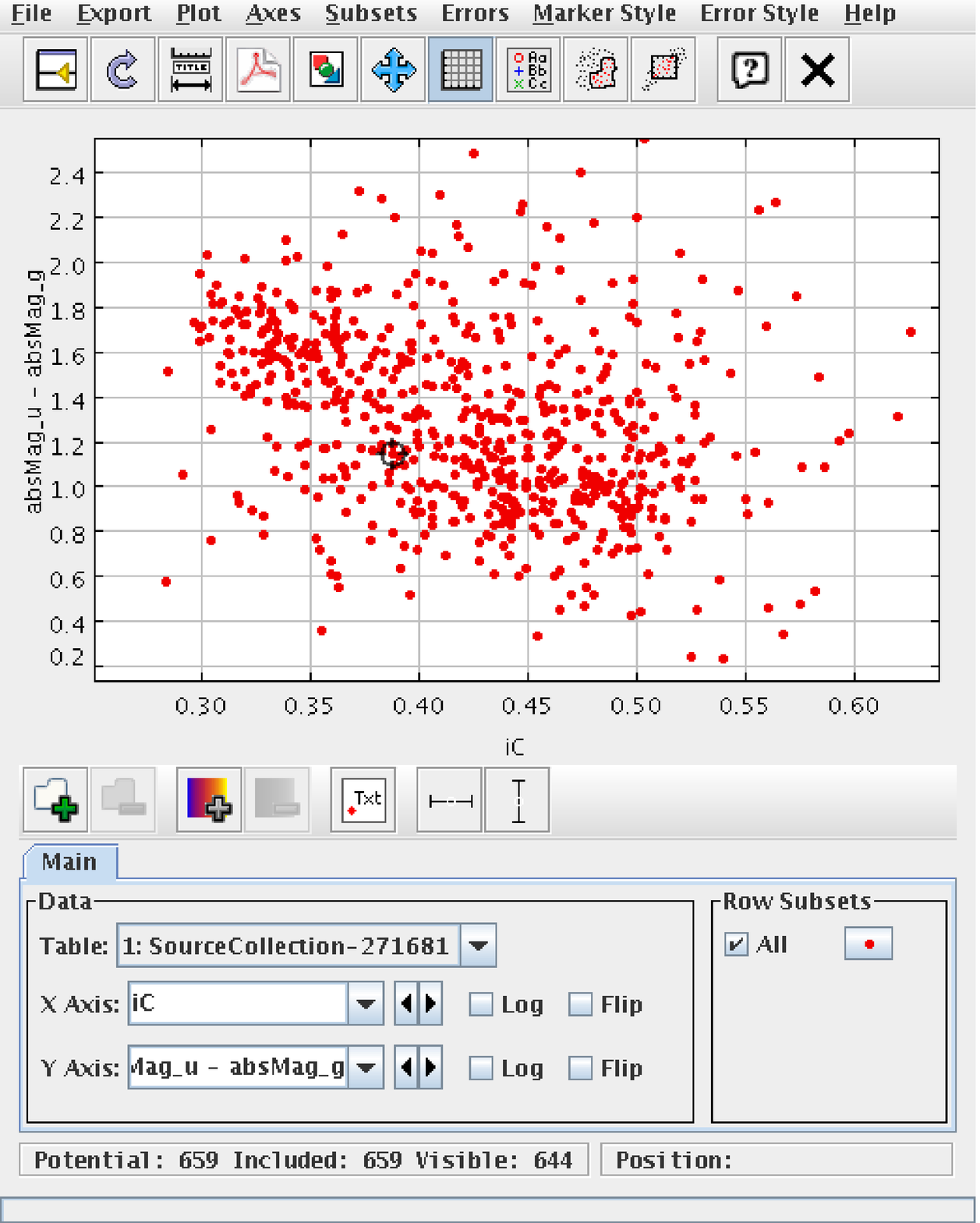}
 \caption{
A Topcat scatter plot showing a color-concentration diagram of the catalog pulled \reffig{fig:scsimplepuller}.
A slight bimodality between red, concentrated, galaxies and blue, extended galaxies can be seen.
}
 \label{fig:qdvtopcat}
\end{figure}

\begin{itemize}
\item
\label{sec:sampsimplepuller}
The \textit{\SimplePuller} (\reffig{fig:scsimplepuller}) represents the most basic way an application can pull catalog data.
Its sole capability is to send a \texttt{\sampcatalogpull} message, it cannot receive messages.
It requires a minimum amount of input (\refsec{sec:sampprotocolqdv}):
\begin{itemize}
 \item An identifier of the base catalog from which the sources are selected.
 \item A list of required attributes.
 \item A query to select the sources.
\end{itemize}
The only knowledge the user needs to have about the information system is how these parameters should be specified.
This service could be built into existing visualization tools quickly.
The demo application uses a web-based interface with the server running locally and relies on other SAMP applications for the actual visualization.

\item
\label{sec:samptreeexplorer}
The \textit{\TreeViewer} (\reffig{fig:sctreeexplorer}) shows how a SAMP application can use the \texttt{\sampcataloginfo} message to give the user more information about the data lineage and derivation of a particular dataset.

This demo application recognizes several of the classes used in \AW and is able to interpret some of their properties. 
The application allows exploration of the dependency graph of a pulled catalog by presenting it as a dot\footnote{\url{http://www.graphviz.org/}} graph.
Clicking on a node sends the \texttt{\sampcataloghighlight} message, allowing interaction with the \aweprompt.

\item
The \textit{\ObjectViewer} (\reffig{fig:scscviewer}) demonstrates how an application can use the object related messages (\texttt{\sampcataloginfo}, \texttt{\sampcatalogchange} and \texttt{\sampcatalogaction}) to influence the properties of process targets and other objects.
It has knowledge about the \AW\ \SourceCollection classes---used to represent astronomical catalogs---and allows many of the actions that can be performed in the \aweprompt\ to be done through the web-based GUI.
\end{itemize}
These applications rely on other SAMP applications for the actual visualization.
For example, Topcat is used in \reffig{fig:qdvtopcat} to visualize the data requested in \reffig{fig:scsimplepuller}.

\section{Example Usage}
The figures depicting the prototype applications show a simple use case of the new message.
First the \SimplePuller (\reffig{fig:scsimplepuller}) is used to request absolute magnitudes and the inverse concentration index for the sources in a specific catalog for which a specific logical expression ($R<300$) holds.
Catalogs in \AW that can be used for data pulling are called \SourceCollections and are identified by an integer, in this case 100511.
Other information systems might use different identifiers.
Attributes are referred to by their name only in \AW.

Subsequently the \TreeViewer (\reffig{fig:sctreeexplorer}) is used to inspect the dependency graph that is proposed to provide the requested catalog.
The \SourceCollection that is responsible for the selection of the sample is highlighted.
The highlighted object is shown in the \ObjectViewer (\reffig{fig:scscviewer}), where the selection criterion is checked and changed if required.

The dependency graph is stored persistently once the scientist has verified that the proposed is suitable for his or her scientific goals, which can be done from the \ObjectViewer as well.
The dependency graph is then optimized automatically before being processed, as described in Paper I.
The catalog data of the last node in the dependency graph is send to \topcat for visualization (\reffig{fig:qdvtopcat}) once it has been processed.

This example shows how a relatively simple request can result in a complex dependency graph.
Nonetheless, this graph can be navigated and changed quickly due to the new SAMP messages.
Furthermore, the catalogs created to fulfill the request are created such that they are most suitable for reuse for later requests and at the same time processed in such a way that minimizes the required calculations.
Newly created catalogs are shared implicitly between collaborating scientists.

Therefore, large datasets can be explored quickly with a high level of flexibility, without requiring the scientist to know details of how the information system handles these large datasets.

\section{SAMP Protocol and Messages}
\label{sec:sampprotocol}
We give the details about SAMP that are necessary to describe our proposed messages and present our extensions and their \AW implementation.

\subsection{SAMP Protocol and Data Types}
SAMP is in principle language-agnostic and is based on abstract interfaces. 
That is, it specifies which functions the HUB and the clients must have in order to send and receive messages, but not the exact protocol that the applications use to call those functions.
The rules which describe how SAMP functions are mapped to the internally used protocol is described in a SAMP \textit{Profile}.
One standard profile based on XML-RPC\footnote{\url{http://www.xmlrpc.com/}} is described in the official documentation, and this is what is used in \AW and in the prototype applications.
XML-RPC is a remote procedure call protocol which uses XML to encode its calls and HTTP as a transport mechanism and is platform independent.

Only three data types are supported in SAMP, because it is language- and even communication-protocol-agnostic:
\begin{itemize}
 \item \texttt{string}: A scalar value consisting of a sequence of ASCII-characters.
 \item \texttt{list}: An ordered array of data items.
 \item \texttt{map}: An unordered associative array with a string as key.
\end{itemize}
Other scalar types have to be mapped to strings, and there is a specification to represent integers, floats and booleans as strings. These data types can be nested to any level: e.g., it is possible to have a map with lists as values.

\label{sec:sampprotocolmessages}
SAMP applications communicate through messages of specific types.
Message types that start with \texttt{samp.} are administrative messages defined by the protocol, the others are defined by application authors.
Clients are supposed to give a general reply with success or failure of a requested operation, even if no response is required.

\subsection{Query Driven Visualization Messages}
\label{sec:sampprotocolqdv}
We designed new SAMP messages and data structures to enable query driven visualization through data pulling mechanisms.
The \texttt{target.object.*} messages assume that the information system uses an object oriented model for science products such as catalogs (\refsec{sec:qdvintro}).
The proposed messages are:
\begin{itemize}
 \item \texttt{\sampcatalogderive}: Create a \sampperscatalog through data pulling. 
Arguments: 
\begin{itemize}
\item \texttt{catalog-id} (string): Identifier of the catalog to select the sources from.
\item \texttt{query} (string): Selection criterion for the sources.
\item \texttt{attributes} (list of strings): Names of the attributes.
\end{itemize}
 \item \texttt{\sampcatalogpull}: Perform the same action as \texttt{\sampcatalogderive} and send the data over SAMP.
Arguments: 
\begin{itemize}
\item \texttt{catalog-id} (string): Identifier of the catalog to select the sources from.
\item \texttt{query} (string): Selection criterion for the sources.
\item \texttt{attributes} (list of strings): Names of the attributes.
\end{itemize}
 \item \texttt{\sampcataloghighlight}: Highlight \asamppersobject.
Arguments: 
\begin{itemize}
\item \texttt{class} (string): Class of the object.
\item \texttt{object-id} (string): Identifier of the object.
\end{itemize}
 \item \texttt{\sampcataloginfo}: Returns a SAMP map with information about \asamppersobject as described below.
Arguments: 
\begin{itemize}
\item \texttt{class} (string): Class of the object.
\item \texttt{object-id} (string): Identifier of the object.
\end{itemize}
 \item \texttt{\sampcatalogchange}: Change a property of \asamppersobject.
Arguments: 
\begin{itemize}
\item \texttt{class} (string): Class of the object.
\item \texttt{object-id} (string): Identifier of the object.
\item \texttt{property-id} (string): Identifier of a property of the object.
\item \texttt{value} (string): New value of the property.
\end{itemize}
 \item \texttt{\sampcatalogaction}: Perform an action related to a \asamppersobject.
Arguments: 
\begin{itemize}
\item \texttt{class} (string): Class of the object.
\item \texttt{object-id} (string): Identifier of the object.
\item \texttt{property-id} (string, optional): Identifier of a property of the object.
\item \texttt{action-id} (string): Identifier of the action.
\end{itemize}
\end{itemize}

\subsection{Query Driven Visualization Data Format}
SAMP data structures are defined to send information about objects between applications.
The structures are designed generic enough that they could be used for any information system.
Information about an object itself, such as the response to an \texttt{\sampcataloginfo} message, is communicated through a map with the following keys:
\begin{itemize}
 \item \texttt{class} (string): The class of the object. A client that has knowledge about the used classes could handle known classes in a special way.
 \item \texttt{id} (string): Identifier this object, unique in combination with the class.
 \item \texttt{status} (string): Indication the processing status of this object (see below).
 \item \texttt{properties} (list of maps): Properties of this object (see below). 
 \item \texttt{actions} (list of maps): Actions that can be performed on this object (see below). 
 \item \texttt{readonly} (boolean): Flag to indicate that the object cannot be modified.
\end{itemize}

\noindent Properties of an object, for example process parameters, are described with a map with the following keys:
\begin{itemize}
 \item \texttt{name} (string): Name of the property, as used by the object.
 \item \texttt{class} (string): The class that the value of the property should have, or a primitive such as `int'.
 \item \texttt{description} (string): A human readable description of the property.
 \item \texttt{value} (string): The used value for the property. This is the \texttt{id} of the object if the property refers to another object.
 \item \texttt{options} (list of maps): Possible values for the property, if applicable.
 \item \texttt{actions} (list of maps): Actions that can be performed on the property.
 \item \texttt{readonly} (boolean): Flag to indicate that the property cannot be modified.
\end{itemize}
\noindent An action that can be performed on an object or property is defined by a map with the following keys:
   \begin{itemize}
     \item \texttt{id} (string): A unique identifier for this action.
     \item \texttt{name} (string): A human presentable name for this action.
   \end{itemize}

\subsection{Query Driven Visualization Object Status}
The status value of an object refers to the processing status of the object.
It can have the following values:
\begin{itemize}
 \item \texttt{\statusok}: The object has been processed, or can be processed while retrieving the result.
 \item \texttt{\statusauto}: The object has to be processed before it can be retrieved. This can be done without user interaction.
 \item \texttt{\statusnew}: This is a non persistent object, which can be processed without user interaction.
 \item \texttt{\statusdepends}: This is a new object, which can be processed only after human intervention. For example, to set a process parameter that has no proper default.
 \item \texttt{\statusnot}: As it is, this object cannot be processed, e.g. because a dependency cannot be fulfilled. The scientist might be able to solve the problem, but whether this is the case is not clear to the information system.
 \item \texttt{\statusunknown}: The status is unknown.
\end{itemize}

\section[SAMP in the Astro-WISE awe-prompt]{SAMP in the \AW\ \aweprompt}
\label{sec:sampastrowise}
The \AW\ \aweprompt is an interactive Python prompt that forms the primary user interface to \AW.
We developed a Python module for \AW to use SAMP from the \aweprompt.
This allows an astronomer to combine the large scale data handling from \AW\ with the visualizations from other \samp applications.

This section is most interesting for readers already familiar with \AW.
All relevant terms are introduced briefly for readers new to \AW.
The SAMP-related functionality that is not query driven visualization specific, is included as well for completeness.

\subsection{SAMP Classes and metadata}
The SAMP module is split up in two classes, a stand-alone Python SAMP client and a derived class with \AW specific functionality:

\begin{itemize}
\item \textbf{SampProxy}:
An instance of the \textit{SampProxy} class is a basic SAMP client.
This class contains all SAMP code that is not \AW specific, and can therefore be used by other Python applications as well.

\item \textbf{Samp}:
The \textit{Samp} class is derived from SampProxy and contains all \AW specific code.
The metadata that the class declares to the HUB ---as stored in its \texttt{metadata} property--- is:
\begin{verbatim}
author.email             buddel@astro.rug.nl
author.name              Hugo Buddelmeijer
home.page                http://www.astro-wise.org
author.affiliation       
    Kapteyn Astronomical Institute, Groningen
samp.name                Astro-WISE
samp.description.html    <p>Astro-WISE</p>
samp.documentation.url   http://www.astro-wise.org
samp.icon.url            
    http://www.astro-wise.org/pics/logo-samp-astrowise.png
samp.description.text    Astro-WISE.
\end{verbatim}

\end{itemize}

\subsection{Sending Data}
All \AW objects that represent catalog or image data can be send over SAMP, using the \texttt{table.load.votable} and \texttt{image.load.fits} messages respectively.

Source catalogs that can be used for data pulling are called \textit{Source Collections} (Paper I).
There are different Source Collection classes, depending on the operation used to create the catalog.
For example, an \textit{Attribute Calculator} Source Collection is used to calculate new attributes (parameters) of sources.
Other catalog related \AW classes that can be send over SAMP are: the \textit{SourceList} which is primarily used to derive parameters directly from images, the non-persistent \textit{TableConverter} to manipulate tabular data in Python and the \textit{PhotSrcCatalog} used for photometric calibration.

Image data in \AW is handled by various \textit{Frame} classes
These are beyond the scope of this paper, because its focus is on catalog data.

\subsection{Catalog Interaction}
The SAMP client supports sending and receiving of both the \texttt{tab\-le.\-high\-light.\-row} and \texttt{tab\-le.\-se\-lect.\-row\-List} messages.
Sources can be highlighted either by their SAMP row id, or through their \AW identifiers.

A SourceList has a \texttt{SLID} as identifier, and sources in a SourceList are labeled with a \texttt{SID}.
The \texttt{SLID}-\texttt{SID} combination uniquely identifies a source.
A Source Collection has a \texttt{SCID} as identifier and can contain sources from multiple SourceLists.

\subsection{Query Driven Visualization Data Structures}
\label{sec:sampaweqdvdata}
Only Source Collection (Paper I) instances can currently be exported over SAMP through the \texttt{\sampcataloginfo} message type. The following properties are send as a reply to such a message:
\begin{itemize}
 \item 
All persistent properties that do not relate to data caching.
References to other \AW objects are exported as the unique identifier of the object.
 \item Process parameters of Attribute Calculators (Paper I) are exported as if they are regular properties.
 \item The names of the attributes are exported as the \texttt{attribute|\%i} properties, where the \texttt{\%i} are consecutive integers. The \texttt{SCID}s of the Source Collections that the attributes originate from are exported as the \texttt{origin|\%i} properties.
\end{itemize}
The actions that can be performed on \sourcecollections are:
\begin{itemize}
 \item \texttt{commit}: Commits a transient Source Collection.
 \item \texttt{copy}: Creates a copy of a Source Collection.
 \item \texttt{make}: Process the \sourcecollection. The exact composition of sources and values of the attributes are determined.
 \item \texttt{send}: Broadcasts the catalog data corresponding to a \sourcecollection over SAMP.
\end{itemize}
Only the \texttt{attribute|\%i} properties have an action:
\begin{itemize}
 \item \texttt{search}: Search for \SourceCollections that could be used as a dependency to provide the attribute. These will be listed in the \texttt{options} of the property.
\end{itemize}

\subsection{Receiving Query Driven Visualization Messages}
The query driven visualization messages from \refsec{sec:sampprotocolmessages} are supported, but only with respect to \SourceCollections.
The actual data pulling is performed with the non-persistent \textit{Source Collection Tree} class.
The parameters of the messages are interpreted as follows:
\begin{itemize}
 \item \texttt{catalog-id}: The \texttt{SCID} of a Source Collection.
 \item \texttt{query}: An Oracle SQL \texttt{WHERE} clause, with attributes in double quotes.
 \item \texttt{attributes}: A list of attribute names.
 \item \texttt{class}: The name of an \AW class. Only Source Collection classes are supported at the moment.
 \item \texttt{object-id}: The \texttt{SCID} of a Source Collection.
 \item \texttt{property-id}: The name of a property. These are either persistent properties as stored in the database, or transient properties that are derived on the fly.
 \item \texttt{action-id}: The identifier of an action that can be performed on an object or property, as defined by the instance itself.
\end{itemize}
The query driven visualization messages are handled as follows:
\begin{itemize}
 \item \texttt{\sampcatalogderive}: The \awfunction{derive} of a Source Collection Tree instance is called to derive a new Source Collection from the specified one.
 \item \texttt{\sampcatalogpull}: Performs the same action as \texttt{\sampcatalogderive} after which the resulting Source Collection is processed and broadcasted.
 \item \texttt{\sampcataloghighlight}: Stores a reference to the highlighted Source Collection as a member of the SAMP instance.
 \item \texttt{\sampcataloginfo}: Returns information about a \SourceCollection.
 \item \texttt{\sampcatalogchange}: Change a property of a \SourceCollection, either directly by the SAMP instance, or by the object itself.
 \item \texttt{\sampcatalogaction}: Perform an action related to a \SourceCollection, either directly by the SAMP instance or by the object itself.
\end{itemize}

\section{Conclusions}
In this paper we see query driven visualization as an extension of data pulling, with a focus on catalog data.
This allows scientists to discover existing datasets and create new datasets by requesting data directly from within the visualization.
New datasets are automatically created in such a way that they are most suitable for reuse in future requests, preventing duplications of data.
The subsequent processing of the datasets is limited to those parts that are necessary to create the data for the requested visualization, achieving implicit scalability.

Requesting existing data and creating new data is done through the same process, because data is found and processed automatically.
The same mechanisms ensure that scientists have control over the methods and parameters that are used to process their data, achieving flexibility.
This allows a high level of abstraction in the interoperation between software, because requests for data can be done in a conceptual way.

The Simple Application Messaging Protocol is an excellent mechanism to provide such a layer abstraction and we proposed new message types to perform query driven visualization.
Support for these messages is implemented within \AW and several prototype applications.

Query driven visualization allows scientists to interact with their data in a conceptual way and allows them to focus on \textit{what} they want to do with the data, because \textit{how} the processing is performed and \textit{where} the data is stored is implicitly taken care of.
Current wide field surveys such as KIDS will produces such large datasets that this automation of administration and implicit scalability is essential.
Therefore, query driven data visualization is not only a bright possible future, but perhaps even an inevitable one.

\begin{acknowledgements}
This research is part of the project ``Astrovis'', research program STARE
(STAR E-Science), funded by the Dutch National Science Foundation (NWO),
project no.\ 643.200.501.
\end{acknowledgements}

\end{document}